# NONLINEAR PREDICTIVE MODELS COMPUTATION IN ADPCM SCHEMES[1]

*Marcos Faundez-Zanuy*

Escola Universitària Politècnica de Mataró

Avda. Puig i Cadafalch 101-111, E-08303 Mataró (BARCELONA)

tel:(34) 93 757 44 04 fax:+(34) 93 757 05 24

e-mail: faundez@eupmt.es http://www.eupmt.es/veu

**ABSTRACT**

Recently several papers have been published on nonlinear prediction applied to speech coding. At ICASSP'98 we presented a system based on an ADPCM scheme with a nonlinear predictor based on a neural net. The most critical parameter was the training procedure in order to achieve good generalization capability and robustness against mismatch between training and testing conditions. In this paper, we propose several new approaches that improve the performance of the original system in up to 1.2dB of SEGSNR (using bayesian regularization). The variance of the SEGSNR between frames is also minimized, so the new scheme produces a more stable quality of the output.

## 1. INTRODUCTION

In our work we use the neural nets with sigmoid transfer function in the hidden layers as a nonlinear predictor. This predictor replaces the LPC predictor in order to obtain an ADPCM scheme with nonlinear prediction.

Classical ADPCM waveform coders compute the predictor coefficients in one of two ways:

a) backward adaptation: The coefficients are computed over the previous frame. Thus, it is not needed to transmit the coefficients of the predictor, because the receiver has already decoded the previous frame and can obtain the same set of coefficients.

b) forward adaptation: The coefficients are computed over the same frame to be encoded. Thus, the coefficients must be quantized and transmitted to the receiver. In [1] we found that the SEGSNR of forward schemes with unquantized coefficients is similar to the classical LPC approach using one quantization bit less per sample. On the other hand with this scheme the mismatch between training and testing phases is smaller than in the previous case, so the training procedure is not as critical as in backward schemes, and the SEGSNR are greater.

In [2] we found that the quantization of the predictor coefficients (forward scheme) is not a trivial question, so in this paper we propose a new training approach of the neural net, in order to improve the performance of the backward scheme. The goal is to obtain similar results with forward (unquantized coefficients) and backward ADPCM nonlinear prediction schemes. On the other hand, this paper also lets to establish several conclusions about how must be trained a neural net in order to achieve good generalization capability.

### 1.1 About nonlinear prediction

Computational issues and complexity of ADPCM with nonlinear prediction was addressed in [3]. Although the computational burden of the proposed system is near 30 times greater than the linear scheme, and other coding methods exist in terms of quality vs bit rate as higher complexity is allowed, some of the best results in terms of optimising both quality and bit rate are obtained from codec structures that contain some form of nonlinearity. Analysis-by-synthesis coders fall into this category. For example, in CELP coders the closed-loop selection of the vector from the codebook can be seen as a data-dependent nonlinear mechanism [4]. Thus, the goal of this paper is to strengthen the knowledge on the bahaviour of a nonlinear predictor speech coder, rather than to propose a "state of the art speech coder". It is important to take into account that the modern speech coders were possible as a evolution of the classical waveform coders, rather than proposing previously an analysis-by-synthesis coder.

## 2. NNET WEIGHT'S COMPUTATION

In our previous work we fixed the structure of the neural net to 10 inputs, 2 neurons in the hidden layer, and one output. The selected training algorithm was the Levenberg-Marquardt, that computes the approximate Hessian matrix, because it is faster and achieves better results than the classical backpropagation algorithm. We also apply a multi-start algorithm with five random initializations for each neural net. Our experiments revealed that the most critical parameter was the number of epochs. The optimal number of epochs is different for each frame [5]. A variable number of epochs for each frame would imply the transmission of this number to the receiver, so we done a statistical study and determined that an optimal (in average) number of epochs was 6. The main problem is that a high number of epochs implies a lose of the generalization capability, and a small number a poor learning.

[1] This work has been supported by the CICYT TIC97-1001-C02-02

In this paper we propose several schemes that improve the generalization capability of the neural net and/or imply an easy determination of the optimal number of epochs without implying the transmission of this parameter.

**2.1 Proposed new schemes**

The propositions of this paper include:

*a) The use of regularization.*

The regularization involves modifying the performance function, which is normally chosen to be the sum of squares of the network errors on the training set. This technique minimizes the effect of overtraining, so more epochs can be done. The classical mean square error function (mse) is replaced by the mean square error regularized function msereg:

$$msereg = \gamma mse + (1-\gamma)\frac{1}{n}\sum_{j=1}^{n} w_j^2$$

where the last term is proportional to the modulus of the weights of the neural net, and is $\gamma$ the performance ratio. In our simulations we have used $\gamma=0.9$. Using this performance function will cause the network to have smaller weights and biases, and this will force the network response to be smoother and less likely to overfit. Figure 1 shows the evolution of the SEGSNR vs $\gamma$, for 6 and 50 epochs. This plot has been obtained with one speaker.

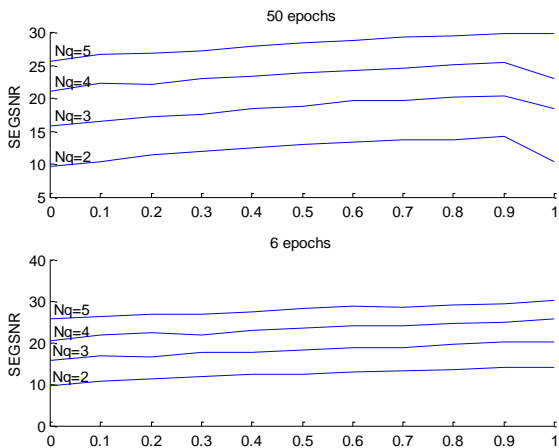

*Figure I: SEGSNR for several performance ratios $\gamma$*

*b) Early stopping with validation*

Early stopping is a technique based on dividing the data into three subsets. The first subset is the training set used for computing the gradient and updating the network weights and biases. The second one is the validation set. The error on the validation set is monitorized during the training process. The validation error will normally decrease during the initial phase of training, as does the training set error. However, when the network begins to overfit the data, the error of the validation set will typically begin to rise. When the validation error increases for a specified number of iterations, the training is stopped, and the weights and biases at the minimum of the validation error are chosen. In this paper we propose to select the training, validation and testing sets as $frame_{k-1}$, $frame_k$, $frame_{k+1}$ respectively (see figure 2).

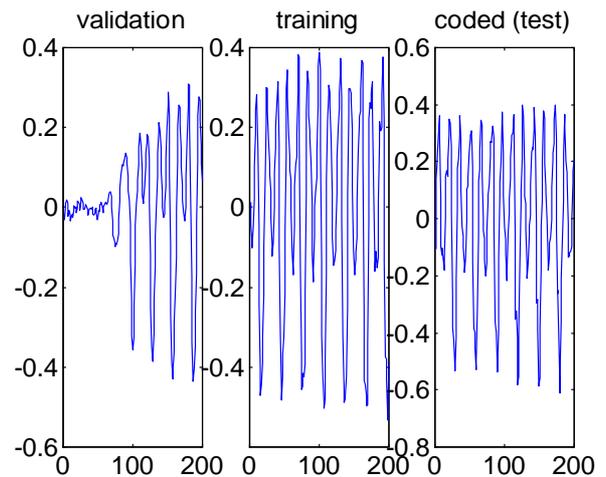

*Figure 2 validation, training and test sets*

Obviously this is a good approach for stationary portions of the signal, which are the most frequent in real speech signals. With this scheme the delay of the encoder is increased in one frame.

*c) Bayesian regularization.*

The approach a) has the problem of estimating the regularization parameter $\gamma$. If it is too large then there will be overfitting, and if it is very small, the neural net will not fit the training samples properly. In order to determine the optimal $\gamma$ in an automated fashion combined with the Levenberg-Marquardt algorithm, we use [6] and [7].

*d) Validation and bayesian regularization.*

Validation and regularization of sections b) and c) are used both together.

*e) Committee of neural nets [8]*

Using a multi-start approach, several neural nets are trained for each frame, and one of them is chosen. We have chosen the neural net that yields the smaller training error, but this criterion does not imply the best performance over the test frame. The main problem with a multi-start algorithm is that it is not possible to choose in advance which is the best neural net, because this imply to test all the nets over the frame we want to encode, and to transmit the index of the selected net to the receiver. Instead of using a multi-start algorithm, it is possible to use all the trained networks, combining their outputs. It is important to see that most of the time is spent in computing the weights of the neural nets. Thus, the computational burden of a committee of neural nets is nearly the same of the multi-start algorithm. The structure is shown in figure 3. Several combinations of the neural net outputs are possible. We have studied the following:

$$\tilde{x}[n] = mean\{\tilde{x}_M[n],\ldots,\tilde{x}_1[n]\}$$
$$\tilde{x}[n] = median\{\tilde{x}_M[n],\ldots,\tilde{x}_1[n]\}$$

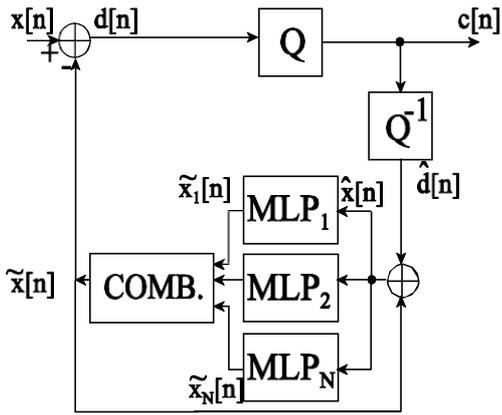

Figure 3: ADPCMB with a committee of nnets. COMB means combination

## 3. RESULTS AND CONCLUSIONS.

The results have been obtained with a database of 8 speakers (4 male & 4 female) sampled at 8kHz, with a frame length of 200 samples, and ADPCM backward scheme. Table 1 shows the results of the classical LPC-10 (same predictor order) and LPC-25 (same number of coefficients). Nq is the number of quantization bits (i.e. Nq=2 implies an ADPCM scheme at 16Kbps). Table 2 is an example of the obtained results with the nonlinear prediction and the proposed training algorithms.

We have used the following notation:
- L-M: Levenberg-Marquardt.
- B-R: Bayesian Regularization.
- V: Validation (proposition *b* of section 2.1).
- Cmean: Committee (proposition *e* of section 2.1) with mean{} combination.
- Cmedian: Committee (proposition *e* of section 2.1) with median{} combination.
- Nq: Number of quantization bits for the adaptive quantizer (=number of bits per sample).

|        | Nq=2  | Nq=3  | Nq=4  | Nq=5  |
|--------|-------|-------|-------|-------|
| LPC-10 | 14.28 | 19.95 | 24.74 | 29.45 |
| LPC-25 | 14.44 | 20.18 | 24.86 | 29.34 |

Table I: SEGSNR of ADPCM with the classical LPC predictor

The most relevant conclusions are:
1. Without regularization, overtraining can reduce the SEGSNR up to 3 dB and increase the variance of the SEGSNR.
2. The use of regularization lets to increase the number of epochs without overfitting the training samples.
3. The use of regularization decreases the variance of the SEGSNR between frames. Thus, the quality of the reconstructed signal has less fluctuation between frames. This is, perhaps, the most relevant achievement of the proposed system.
4. Automatic selection of the regularization parameter $\gamma$ (Bayesian regularization) must be used in order to improve the SEGSNR of the decoded speech signal. In this case, an increase of 1.2dB is achieved for Nq=2, and 0.8 dB in the other cases. On the other hand, for Nq=2 & 3 it is better to fix the number of epochs to 6 and for Nq=4 & 5 to 50 epochs. This is because the mismatch between training and testing is greater for small number of quantization bits (the quantization error degrades more seriously the predictor input signal).
5. The use of validation produces minor effects, and not always improves the SEGSNR or the variance of the SEGSNR.
6. The combination using the *median{}* function yields better results than *mean{}* function. In [3] we shown a high variance for several nets (with different random initialization) over the same training frame. This variance was measured in terms of prediction gain (Gp). Thus, some random initializations can get stuck in a local minimum. This will result in a bad prediction, that can be removed with a median{} filter and a committee of several networks, because for most of the speech frames, the number of "good" initializations is greater than the number of "bad" random initializations.

| Training algorithm | performance function | epoch | Nq=2 | | Nq=3 | | Nq=4 | | Nq=5 | |
|---|---|---|---|---|---|---|---|---|---|---|
| | | | SEG SNR | std | SEG SNR | std | SEG SNR | std | SEG SNR | std |
| L-M | mse | 6 | 13.7 | 6 | 20.4 | 7.2 | 25.5 | 7.5 | 30.4 | 7.7 |
| L-M,Cmean | mse | 6 | 14.2 | 5.3 | 20.6 | 6.2 | 25.9 | 6.5 | 30.4 | 6.8 |
| L-M,Cmedian | mse | 6 | 14.5 | 5.4 | 21 | 6.2 | 26.2 | 6.7 | 31 | 7 |
| L-M | mse | 50 | 10.3 | 10.6 | 18.1 | 11.1 | 23.3 | 11.1 | 28.2 | 12.1 |
| L-M,Cmean | mse | 50 | 11.7 | 10.6 | 18.7 | 10.9 | 24.1 | 10.8 | 29.4 | 10.9 |
| L-M,Cmedian | mse | 50 | 13.7 | 7 | 20.5 | 7.7 | 25.8 | 7.9 | 30.9 | 7.9 |
| L-M | msereg | 6 | 13.6 | 5.5 | 19.15 | 6.9 | 24.3 | 7 | 29.1 | 6.9 |
| L-M,Cmean | msereg | 6 | 13.9 | 5.3 | 19.8 | 5.8 | 25 | 6.4 | 29.7 | 6.8 |
| L-M,Cmedian | msereg | 6 | 14.2 | 5.1 | 20.2 | 6 | 25.3 | 6.4 | 30 | 6.8 |
| L-M | msereg | 50 | 13.8 | 5.3 | 19.5 | 7.6 | 24.6 | 7.1 | 29.3 | 7.6 |
| L-M, Cmean | msereg | 50 | 14.5 | 5.7 | 20.4 | 6.3 | 25.4 | 6.7 | 30.3 | 6.9 |
| L-M, Cmedian | msereg | 50 | 14.4 | 5.4 | 20.5 | 6.2 | 25.6 | 6.6 | 30.3 | 6.9 |
| B-R | msereg | 6 | 14.9 | 5.3 | 21.2 | 6.3 | 26.2 | 6.6 | 30.8 | 7 |
| B-R, Cmean | msereg | 6 | 14.5 | 4.9 | 20.5 | 5.9 | 25.6 | 6.5 | 30.5 | 6.8 |
| B-R, Cmedian | msereg | 6 | 14.9 | 5.2 | 21 | 6 | 25.9 | 6.5 | 30.7 | 6.9 |
| B-R | msereg | 50 | 14.2 | 5.7 | 21 | 6.3 | 26.3 | 6.7 | 31.1 | 7.1 |
| B-R, Cmean | msereg | 50 | 14.6 | 5.6 | 21.3 | 6.4 | 26.5 | 6.8 | 31.4 | 7.2 |
| B-R, Cmedian | msereg | 50 | 14.3 | 5.5 | 21.1 | 6.2 | 26.4 | 6.8 | 31.2 | 7.1 |
| L-M, V | mse | 50 | 13.2 | 7.4 | 19.8 | 8 | 25 | 8.1 | 30.3 | 8.1 |
| L-M, V, Cmean | mse | 50 | 14 | 7 | 20.5 | 7.8 | 25.7 | 7.9 | 30.6 | 8.1 |
| L-M, V, Cmedian | mse | 50 | 14.5 | 6.3 | 21.1 | 6.9 | 26.1 | 7.2 | 31 | 7.3 |
| L-M, V | msereg | 50 | 13.2 | 6.3 | 19.3 | 7.8 | 24.2 | 7 | 29.2 | 7 |
| L-M, V, Cmean | msereg | 50 | 13.9 | 5.5 | 19.9 | 6.5 | 24.8 | 7.4 | 29.5 | 7.5 |
| L-M, V, Cmedian | msereg | 50 | 14.3 | 6.1 | 20.2 | 6.8 | 25.2 | 6.8 | 30 | 7.2 |
| B-R, V | msereg | 50 | 14 | 6.5 | 20.7 | 7.3 | 26 | 7.2 | 31.1 | 7.6 |
| B-R, V, Cmean | msereg | 50 | 14.4 | 5.8 | 21.1 | 6.7 | 26.4 | 7.3 | 31 | 7.7 |
| B-R, V, Cmedian | msereg | 50 | 14.2 | 6.1 | 20.9 | 6.8 | 26.4 | 6.9 | 31.2 | 7.2 |

*Table 2: SEGSNR and standard deviation of ADPCMB NL prediction with the proposed training algorithms*